\documentclass[
]{jss}

\usepackage{thumbpdf,lmodern}

\usepackage[utf8]{inputenc}

\author{
Boyi Guo\\University of Alabama at Birmingham \And Nengjun
Yi\\University of Alabama at Birmingham
}
\title{The R Package \pkg{BHAM}: Fast and Scalable Bayesian Hierarchical
Additive Model for High-dimensional Data}

\Plainauthor{Boyi Guo, Nengjun Yi}
\Plaintitle{The R Package BHAM: Fast and Scalable Bayesian Hierarchical
Additive Model for High-dimensional Data}
\Shorttitle{\pkg{BHAM}: Bayesian Hierarchical Additive Model}

\Abstract{
BHAM is a freely avaible R pakcage that implments Bayesian hierarchical
additive models for high-dimensional clinical and genomic data. The
package includes functions that generalized additive model, and Cox
additive model with the spike-and-slab LASSO prior. These functions
implement scalable and stable algorithms to estimate parameters. BHAM
also provides utility functions to construct additive models in high
dimensional settings, select optimal models, summarize bi-level variable
selection results, and visualize nonlinear effects. The package can
facilitate flexible modeling of large-scale molecular data,
i.e.~detecting susceptible variables and infering disease diagnostic and
prognostic. In this article, we describe the models, algorithms and
related features implemented in BHAM. The package is freely available
via the public GitHub repository \url{https://github.com/boyiguo1/BHAM}.
}

\Plainkeywords{additive model, spike-and-slab LASSO, scalable}

\Address{
    Boyi Guo\\
    University of Alabama at Birmingham\\
    1665 University Blvd\\
Birmingham, AL 35294-0002 USA\\
  E-mail: \email{boyiguo1@uab.edu}\\
  URL: \url{http://boyiguo1.github.io}\\~\\
      Nengjun Yi\\
    University of Alabama at Birmingham\\
    1665 University Blvd\\
Birmingham, AL 35294-0002 USA\\
  E-mail: \email{nyi@uab.edu}\\

  }

\usepackage{amsmath}

\begin{document}

\newcommand{\pr}{\text{Pr}}
\newcommand{\bs}[1]{\boldsymbol{#1}}
\newcommand{\tp}{*}
\newcommand{\simiid}{\overset{\text{iid}}{\sim}}

\section{Introduction}

High-dimensional statistics has been an indispensable area of research
for its high impact in molecular and clinical data analysis. In recent
years, there are continuous efforts to make high-dimensional models more
flexible and interpretable, aiming to capture more complex signals. One
particular family of such flexible and interpretable models is the
additive models where predictors are included in the model as an
additive function. These high-dimensional additive models serve for two
purposes: variable selection and outcome prediction. In high-dimensional
statistics, it is common to assume there is only a small subset of
predictors that have effects on the outcome, also known as the signal
sparsity assumption. \citep{buhlmann2011} The high-dimensional additive
models select not only the predictors who have linear associations with
the outcome, but also those who inform the outcome prediction with
nonlinearity. As a result, they provide more flexible effect modeling
and improve prediction accuracy compared to high-dimensional linear
models.

There are many proposals on high-dimensional additive models. The main
idea of these proposals focus on the application of grouped sparse
penalties, for example, group LASSO penalty
\citep{ravikumar2009, huang2010} and group SCAD penalty
\citep{wang2007, xue2009}, on the coefficients of additive functions.
There methods are developed primarily for variable selection and may
provide inaccurate estimation of the underlying functions due to the
excess shrinkage of the sparsity penalty \citep{scheipl2013}. Thus, the
prediction performance will be affected. In addition, these methods take
an ``all-in-all-out'' approach for variable selection, and fails to
answer if the underlying signals are linear or nonlinear. To address
these shortcomings, Guo and his colleagues proposed the two-part
spike-and-slab LASSO prior for generalized additive models
\citep{guo2022_GAM} and additive Cox proportional hazards models
\citep{guo2022_Cox}. Instead of using the computationally prohibitive
Markov chain Monte Carlo approximations, optimization-based
EM-Coordinate Descent algorithms are developed for model fitting. Monte
Carlo studies and real data analysis demonstrate improved prediction and
computation performances compare to the state-of-the-art additive
models.

In this article, we introduce an R package \texttt{BHAM} that implements
the spike-and-slab LASSO additive models and computationally efficient
algorithms. Notably, \texttt{BHAM} provides functions for setting up and
fitting various spike-and-slab LASSO additive models, including
generalized additive models for various continuous and discrete outcomes
and Cox proportional hazards models for censored survival outcomes. The
specification of additive functions follows a popular syntax implemented
in \texttt{mgcv} \citep{r_mgcv}. We provide a parser function that
translates high-dimensional predictors names and their corresponding
additive functions to model formulas, rendering convenience to model
large datasets with hundreds and thousands of predictors. Other
ancillary functions include cross-validation, model summary, and effect
visualization. Our objective with \texttt{BHAM} is to offer a friendly
user experience that emphasizes statistical validity, computational
scalability and utility flexibility for high-dimension additive models.

There are other R packages that facilitate flexible modeling of complex
signals via additive models for high-dimensional data analysis. The R
package \texttt{COSSO} \citep{r_cosso} implements smoothing spline ANOVA
models with the component selection and smoothing operator to analyze
generalized and survival outcomes. The packages \texttt{spikeSlabGAM}
\citep{r_spikeSlabGAM} and \texttt{sparseGAM} \citep{r_sparseGAM} fits
generalized additive regression models with spike-and-slab and
spike-and-slab LASSO priors respectively; nevertheless, both packages
does not offer analytic support to model time-to-event outcome. The
package \texttt{tfCox} \citep{r_tfCox} implements additive Cox
proportional hazards models with trend filtering. \cite{scheipl2013}
summarized some other scripts or packages to fit additive models
published before 2013, including \texttt{spam} \citep{ravikumar2009},
\texttt{hgam} \citep{meier2009} and \texttt{hypergsplines}
\citep{bove2011}; unfortunately, these tools are hardly available now
due to maintenance issues. One inconvenience shared by the packages is
the limited ability to customize additive functions due to the
difficulty to formulate the high-dimensional model. In the proposed
\texttt{BHAM}, we address this challenge by providing an interface that
parses a data frame of spline function specification to model formula,
and hence provide greater flexibility compared to previous packages.

The remainder of this paper is as follows. In Section 2, we briefly
describe the spike-and-slab LASSO prior of smooth functions and the
computationally efficient EM-Coordinate Descent algorithm for model
fitting. Section 3 demonstrates the analytic pipeline to analyze
high-dimensional data with the R package \texttt{BHAM}. We deliver the
conclusion in Section 4. Fore more details and examples about
\texttt{BHAM}, we encourage the readers to visit
\url{https://boyiguo1.github.io/BHAM/}.

\section{Models and Algorithms}

In this section, we describe the Bayesian hierarchical additive model
that \texttt{BHAM} implements. The key idea is to impose the two-part
spike-and-slab LASSO prior \cite{guo2022_GAM} on each additive function
in generalized models or Cox proportional hazards models. For the
additive function \(B_j(X_j)\) of the \(j\)th variable, the proposed
two-part spike-and-slab LASSO prior consists of a spike-and-slab LASSO
prior for the linear space coefficient \(\beta_j\) and a modified group
spike-and-slab LASSO prior for the nonlinear space coefficients
\(\beta_{jk}^*, k = 1,..., K_j\), \begin{align}
  \beta_{j} | \gamma_{j},s_0,s_1 &\sim (1-\gamma_{j}) DE(0, s_0) + \gamma_{j} DE(0, s_1)\nonumber \\
  \beta^*_{jk} | \gamma^*_{j},s_0,s_1 &\overset{\text{iid}}{\sim}(1-\gamma_{j}^*) DE(0, s_0) + \gamma_{j}^*DE(0, s_1), k=1,\dots, K_j.
\end{align} To note, the model matrix of each additive function
undergoes a reparameterization process in advance, which eigendecomposes
the smoothing penalty matrix to isolate the linear and nonlinear spaces
of the additive function \citep{wood2017}. The reparameterization
greatly reduces the complexity to formulate the sparsity-smoothness
penalty \citep{meier2009} and allows different shrinkage on the two
spaces. The shrinkage on the linear space manages the variable
selection, while the shrinkage on the nonlinear space emphasizes the
adequate smoothing of nonlinear effect interpolation. In addition, the
isolation of linear and nonlinear spaces motivates the bi-level
functional selection, i.e.~the selection of additive functions and the
selection of nonlinear effects. In the proposed prior, each additive
function has two indicators \(\gamma_{j}\) and \(\gamma^*_{j}\),
controlling the linear and nonlinear component selection. Effect
hierarchy was implemented via the conditional priors of \(\gamma^*_{j}\)
to ensure the the linear component is more likely to be selected than
the nonlinear components. \begin{align}
&\gamma_{j} | \theta_j \sim Bin(1, \theta_j) & &
&\gamma_{j}^*| \gamma_{j}, \theta_j \sim Bin(1, \gamma_{j}\theta_j).
\end{align} We further impose a beta prior on the inclusion probability
parameter \(\theta_j\) to allow locally adaptive shrinkage. For
simplicity, \(\theta_j\) follows a uniform(0, 1) prior. Compared to
previous spike-and-slab priors \citep{scheipl2012, bai2021} for additive
functions, the proposed prior provides three advantages. First of all,
the proposed prior allows bi-level functional selection instead of an
``all-in-all-out'' approach for variable selection. Secondly, the
proposed prior offers a natural selection procedure by shrinking
unnecessary coefficients to exactly 0, contrasting to soft-thresholding
the inclusion probability. Last but not least, the proposed prior is
easily applicable to model different types of outcomes, including
time-to-event outcomes via Cox proportional hazards models.

To fit the proposed models in a efficient and scalable fashion, we
develop the EM-coordinate descent algorithm. The EM-coordinant descent
algorithm estimates maximum a posteriori of coefficients by optimizing
the log joint posterior density function. The algorithm formulates the
spike-and-slab LASSO prior as a double exponential distribution with a
conditional scale parameter. It further leverages the relationship
between double exponential prior and \(l_1\) penalty and expresses the
log joint posterior density function as the summation of a \(l_1\)
penalized likelihood function ( and \(l_1\) penalized partial likelihood
function for Cox proportional hazards models) and log beta posterior
densities. Because the nuisance parameters \(\boldsymbol{\gamma}\) are
unknown, we instead optimize the conditional expectation of log joint
posterior density function via an EM procedure \citep{dempster1977}. In
each iterations of the EM procedure, we update the expectation of the
log joint posterior density function with respect to the nuisance
parameters, calculate the penalties based on the estimation from
previous iteration, and optimize the penalized likelihood and the
posterior density with coordinate descent algorithm and closed-form
calculation for the coefficients. The process iterates until
convergence. Cross-validation is used to choose the optimal model. We
defer to \cite{guo2022_GAM, guo2022_Cox} for more detail on the GAM
algorithm and the additive Cox model algorithm.

\section{Analytic Pipline Using BHAM}

In this section, we demonstrate how to fit Bayesian hierarchical
additive model with two-part spike-and-slab LASSO prior using the
package \texttt{BHAM}. Specifically, we introduce how to 1) prepare the
high-dimensional design matrix for fitting the proposed model, 2) fit
generalized additive model, 3) tune models and assess model performance,
and 4) visualize the bi-level variable selection.

\subsection{Installation}

To install the latest development version of the \texttt{BHAM} package
from \textbf{GitHub}, type the following command in R console:

\begin{CodeChunk}
\begin{CodeInput}
R> if (!require(devtools)) install.packages("devtools")
R> if(!require(BHAM)) devtools::install_github("boyiguo1/BHAM", build_vignettes = FALSE)
\end{CodeInput}
\end{CodeChunk}

You can also set \texttt{build\_vignettes=TRUE} but this will slow down
the installation drastically (the vignettes can always be accessed
online anytime at
\href{https://boyiguo1.github.io/BHAM/articles}{boyiguo1.github.io/BHAM/articles}).

\subsection{Preliminaries}

We use a simulated data set to demonstrate our package. The data
generating mechanism is motivated by \citet{bai2021} and programmed in
the function \texttt{sim\_Bai}: we assume there are \(p=10\) predictors
where the first four predictors have effects on the outcome (see
functions below), and the rest of predictors don't, i.e
\(B_j(x_j) = 0, j = 5, \dots, p\).

\begin{align*}
B_1(x_1) &= 5 \sin(2\pi x_1) & & B_2(x_2) &= - 4 \cos(2\pi x_2 -0.5)\\
B_3(x_3) & = 6(x_3-0.5) & & B_4(x_4) &= - 5(x_4^2 -0.3)
\end{align*}

Using this data generating mechanism, we simulate two datasets of binary
outcomes with the logit link function from Bernoulli trials. To note,
the function \texttt{sim\_Bai} can also simulate Gaussian and Poisson
outcomes using the same data generating mechanism. The sample sizes of
these two datasets are 500 and 1000 for training and testing
respectively. The following code section creates the training and
testing datasets.

\begin{CodeChunk}
\begin{CodeInput}
R> library(BHAM)
R> set.seed(1) ## simulate some data...
R> n_train <- 500
R> n_test <- 1000
R> p <- 10
R> # Train Data
R> train_dat <- sim_Bai(n_train, p)
R> dat <- train_dat$dat %>% data.frame
R>
R> # Test Data
R> test_tmp <- sim_Bai(n_test, p)
R> test_dat <- test_tmp$dat %>% data.frame
\end{CodeInput}
\end{CodeChunk}

The first ten observations of the training data set look like below.

\begin{CodeChunk}
\begin{CodeOutput}
           x1         x2         x3          x4         x5          x6
1   1.5579537 -1.1346302  0.5205997  0.73911492 -1.8054836 -0.88614959
2  -0.7292970  0.7645571  0.3775619  0.38660873 -0.6780407 -1.92225490
3  -1.5039509  0.5707101 -0.6236588  1.29639717 -0.4733581  1.61970074
4  -0.5667870 -1.3516939 -0.5726105 -0.80355836  1.0274171  0.51926990
5  -2.1044536 -2.0298855  0.3125012 -1.60262567 -0.5973876 -0.05584993
6   0.5307319  0.5904787 -0.7074278  0.93325097  1.1598494  0.69641761
7   1.6176841 -1.4130700  0.5212035  1.80608925 -1.3332269  0.05351568
8   1.1845319  1.6103416  0.4481880 -0.05650363 -0.9257557 -1.31028350
9   1.8763334  1.8404425 -0.5053226  1.88591132 -1.0744951 -2.12306606
10 -0.4557759  1.3682979 -0.2066122  1.57838343 -1.4511165 -0.20807859
           x7          x8          x9        x10 y
1   0.8500435  1.13496509  0.07730312 -0.6264538 0
2  -0.9253130  1.11193185 -0.29686864  0.1836433 1
3   0.8935812 -0.87077763 -1.18324224 -0.8356286 0
4  -0.9410097  0.21073159  0.01129269  1.5952808 0
5   0.5389521  0.06939565  0.99160104  0.3295078 0
6  -0.1819744 -1.66264885  1.59396745 -0.8204684 0
7   0.8917676  0.81083998 -1.37271127  0.4874291 0
8   1.3292082 -1.91234580 -0.24961093  0.7383247 1
9  -0.1034661 -1.24675343  1.15942453  0.5757814 0
10  0.6150646  0.99815445 -1.11422235 -0.3053884 0
\end{CodeOutput}
\end{CodeChunk}

\subsection{Set up Design Matrix of additive functions}

Given the raw data, we would like to translate the additive functions to
the their matrix form. The challenge here is to provide convenient way
to specify the high-dimensional model with enough flexibility to
customize the additive functions. Our solution here is to use a data
frame to accommodate each predictor in the raw data set and allow each
predictor have their spline function specified respectively. There are
three columns for this model specification data frame, including
\texttt{Var} \texttt{Func}, \texttt{Args}. The \texttt{Var} column hosts
the variable name; the \texttt{Func} column hosts the spline function
following the commonly used syntax from \texttt{mgcv}; the \texttt{Args}
column hosts the detail specification of the spline function. The data
frame can be constructed manually for low-dimensional settings and also
be manipulated easily when the number of spline components grows to tens
or hundreds. See the examples below.

\begin{CodeChunk}
\begin{CodeInput}
R> # Low-dimensional setting
R> mgcv_df <- dplyr::tribble(
+   ~Var, ~Func, ~Args,
+   "X1",  "s", "bs='cr', k=5",
+   "X2",  "s", NA,
+   "X3",  "s", "",
+ )
R>
R> # High-dimensional setting
R> mgcv_df <- data.frame(
+   Var = setdiff(names(dat), "y"),
+   Func = "s",
+   Args ="bs='cr', k=7"
+ )
\end{CodeInput}
\end{CodeChunk}

After having the model specification data frame, the next task is to
construct the overall design matrix. We provide a function
\texttt{construct\_smooth\_data} to construct the design matrix for each
predictor according to their spline specification. Then we bind the
design matrices of all spline functions together with a systematic
naming convention. The linear component of each spline function is named
with the suffix \texttt{.null} and the nonlinear components are named
with the suffix \texttt{.pen}. In \texttt{construct\_smooth\_data}, we
take three steps of matrix manipulation via the \texttt{smoothCon} from
the package \texttt{mgcv}: 1) set up linear constraints for
identifiability, 2) eigendecomposition of the smoothing matrix \(S\) to
isolate linear and nonlinear spaces, 3) scaling of the design matrix
such that the coefficients are on the same scale. As we use
\texttt{mgcv::smoothCon} to decode the spline specification, we carry
over the ability to work with user-defined spline functions as long as
it follows \texttt{mgcv} standard.

The \texttt{construct\_smooth\_data} function has two arguments, the
model specification data frame and the raw data. It returns the
finalized design matrix \texttt{data} and the smooth specification
functions \texttt{Smooth} which will later be used to construct the
design matrix of the new datasets for prediction.

\begin{CodeChunk}
\begin{CodeInput}
R> train_sm_dat <- BHAM::construct_smooth_data(mgcv_df, dat)
R> train_smooth <- train_sm_dat$Smooth
R> train_smooth_data <- train_sm_dat$data
\end{CodeInput}
\end{CodeChunk}

\subsection{Fitting the Bayesian Hierarchical model}

With the additive function design matrix constructed, we are ready to
fit the Bayesian hierarchical model with the two-part spike-and-slab
LASSO prior. The model fitting algorithm, implementing the EM-coordinate
descent algorithm, is wrapped in the function \texttt{bamlasso}. The
necessary arguments are \texttt{x} for the design matrix, \texttt{y} for
the outcome, \texttt{family} for the family distribution of the outcome,
and \texttt{group} for the additive functions. We provide a utility
function \texttt{make\_group} to automate the grouping, by organizing
column names from the design matrix. It generates a list of vectors
containing the bases of each additive function. Another important
argument is \texttt{ss}, which is a vector of length 2 for scale
parameters of the spike and slab densities. To recall, the
spike-and-slab LASSO prior can be formulated as the mixture of two
double exponential distributions of mean 0, and hence has two scale
parameters. The argument \texttt{ss} defaults to a spike double
exponential density with scale parameter 0.04, and a slab double
exponential density with scale parameter 0.5. These scale parameters is
a general starting value based on empirical evidence
\citep{Tang2018, Tang2019}.

\begin{CodeChunk}
\begin{CodeInput}
R> bham_mdl <- bamlasso(x = train_smooth_data, y = dat$y,
+                      family = "binomial",
+                      group = make_group(names(train_smooth_data)))
\end{CodeInput}
\end{CodeChunk}

\subsubsection{Tuning via Cross-validation}

With the specified \texttt{ss} argument, the function \texttt{bamlasso}
fit the model. Nevertheless, the fitted model may not be the optimal
model. To select the optimal model, we employ a tuning step via cross
validation, which is implemented in the function \texttt{tune.bgam}. The
main arguments are the previously fitted model where the model data,
additive function specifications are stored, a sequence of spike density
scale parameter \(s_0\), and number of folds. The following example
shows to use five-fold cross validation to examine a vector of \(s_0\)
options, from 0.005 to 0.1 with 0.01 increments. Currently, we don't
consider the examination of the slab density scale parameter \(s_1\) for
computational economy. Previously literature \citep{tang2017, Tang2017a}
shows \(s_1\) has modest impact on the model performance. The tuning
function also allows nested cross-validation by allowing running
multiple cross-validation via \texttt{ncv} and user-specified folds via
\texttt{foldid}.

\begin{CodeChunk}
\begin{CodeInput}
R> s0_seq <- seq(0.005, 0.1, 0.01)
R> cv_res <- tune.bgam(bham_mdl, nfolds = 5, s0= s0_seq, verbose = FALSE)
\end{CodeInput}
\begin{CodeOutput}
Fitting ncv*nfolds = 5 models:
1 2 3 4 5
 Cross-validation time: 0.01 minutes
Fitting ncv*nfolds = 5 models:
1 2 3 4 5
 Cross-validation time: 0.005 minutes
Fitting ncv*nfolds = 5 models:
1 2 3 4 5
 Cross-validation time: 0.005 minutes
Fitting ncv*nfolds = 5 models:
1 2 3 4 5
 Cross-validation time: 0.005 minutes
Fitting ncv*nfolds = 5 models:
1 2 3 4 5
 Cross-validation time: 0.005 minutes
Fitting ncv*nfolds = 5 models:
1 2 3 4 5
 Cross-validation time: 0.005 minutes
Fitting ncv*nfolds = 5 models:
1 2 3 4 5
 Cross-validation time: 0.005 minutes
Fitting ncv*nfolds = 5 models:
1 2 3 4 5
 Cross-validation time: 0.005 minutes
Fitting ncv*nfolds = 5 models:
1 2 3 4 5
 Cross-validation time: 0.005 minutes
Fitting ncv*nfolds = 5 models:
1 2 3 4 5
 Cross-validation time: 0.005 minutes
\end{CodeOutput}
\end{CodeChunk}

The cross-validation tuning function returns different performance
metrics, including deviance, mean squared error, mean absolute error,
area under the curve, misclassifcation for binary outcome, and
concordance statistics for survival outcome. The following shows the
cross-validated performance metrics for the first five values of the
\(s_0\) sequence using out-of-bag samples.

\begin{CodeChunk}
\begin{CodeInput}
R> head(cv_res, 5)
\end{CodeInput}
\begin{CodeOutput}
     s0 deviance   auc   mse   mae misclassification
1 0.005  435.044 0.809 0.141 0.281             0.212
2 0.015  376.082 0.865 0.120 0.253             0.166
3 0.025  352.728 0.883 0.111 0.238             0.148
4 0.035  349.896 0.882 0.110 0.226             0.154
5 0.045  346.670 0.884 0.109 0.223             0.158
\end{CodeOutput}
\end{CodeChunk}

Here we want to caution the reader, if the performance metric varies
monotonically with the candidate \(s_0\) values, it would be better to
examine a broader range of candidate \(s_0\) values, as the sequence
contains a local optimal where the global optimal is not reached yet.
Using some visual aid to examine the \(s_0\) and performance metric
relationship would be more helpful.

\begin{CodeChunk}
\begin{CodeInput}
R> plot(cv_res$s0, cv_res$deviance)
R> lines(cv_res$s0, cv_res$deviance)
\end{CodeInput}


\begin{center}\includegraphics{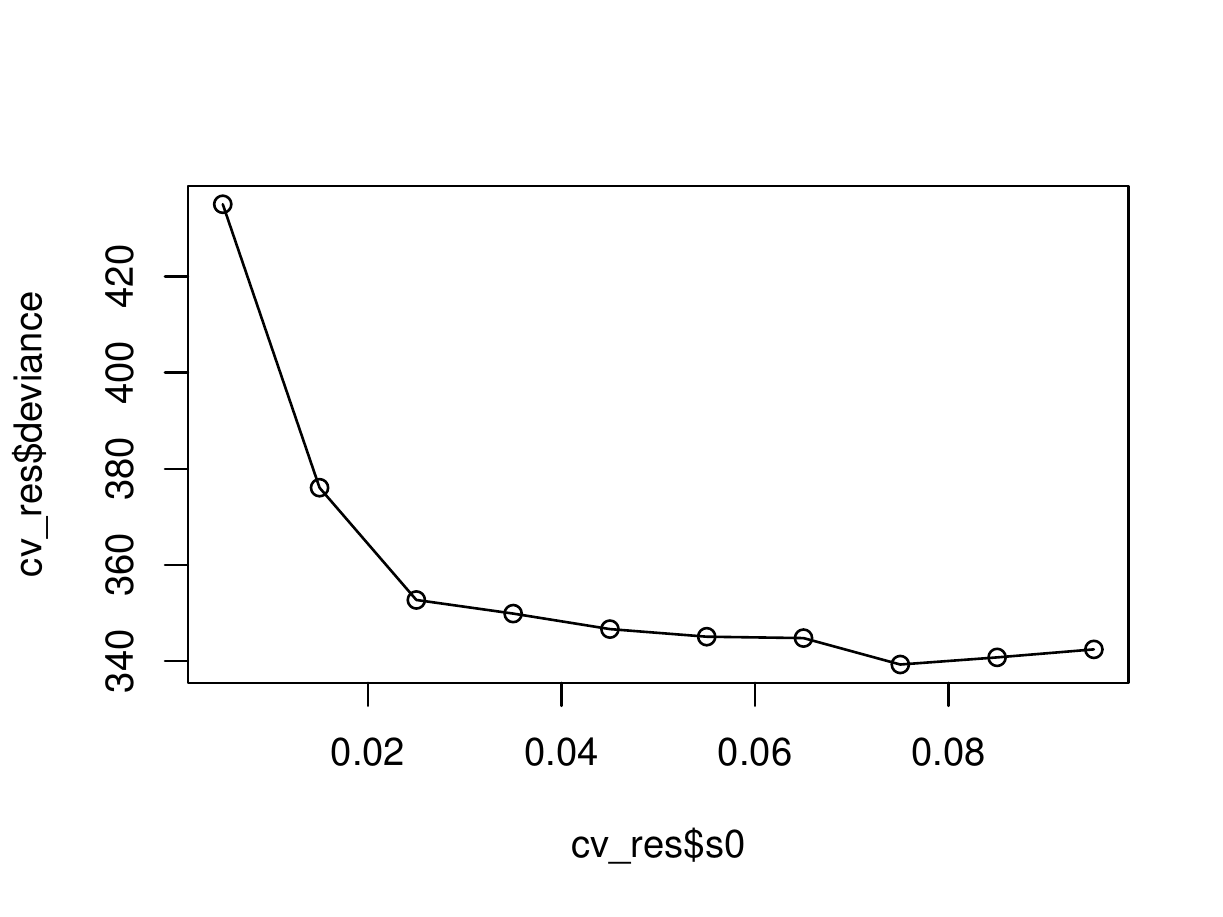} \end{center}

\end{CodeChunk}

With the cross-validation results, we can choose from all the candidate
values of \(s_0\) and select the one with the best performance using the
preferred metrics. For example, we can use the \(s_0\) value that gives
the minimum cross-validated deviance and re-fit the model. Hence, this
would be the optimal model.

\begin{CodeChunk}
\begin{CodeInput}
R> s0_min <- cv_res$s0[which.min(cv_res$deviance)]
R> bham_final <- bamlasso(x = train_smooth_data, y = dat$y,
+                        family = "binomial",
+                        group = make_group(names(train_smooth_data)),
+                        ss = c(s0_min, 0.5))
\end{CodeInput}
\end{CodeChunk}

To note, it is a convention to use some predictive metrics to select the
best performed model among all the candidate values for both predictive
purpose and variable selection purpose. However, previous literature
\citep{Wu2019} shows that when using predictive metrics to select model
for variable selection purpose, the variable selection performance may
not be optimal.

\subsection{Varible Selection and Curve Intropolation}
\subsubsection{Variable Selecrtion}

We provide a function to summarize the variable selection result of a
produced model, namely \texttt{bamlasso\_var\_selection}. The input of
the function is a fitted BHAM model, and the output is a list containing
two components, \texttt{parametric} and \texttt{non-parametric}. The
\texttt{parametrc} component is a vector contains the selected variables
that were fitted in the model in their parameteric form, i.e.~not
specified via additive functions. The \texttt{non-parametric} component
contains a data frame with 3 columns, \texttt{Variable},
\texttt{Linear}, \texttt{Nonlinear}. While \texttt{Variable} column
includes the variable names of selected additive functions,
\texttt{Linear} and \texttt{Nonlinear} columns are logical vectors
indicating if the linear and nonlinear components of additive functions
are included in the model respectively.

\begin{CodeChunk}
\begin{CodeInput}
R> bamlasso_vs_part <- bamlasso_var_selection(bham_final)
\end{CodeInput}
\end{CodeChunk}

Here, we shows the variable selection result from previously tuned
model. Since, the model didn't include any variables in their parametric
form. Hence, the \texttt{parametric} is an empty vector. Meanwhile, the
\texttt{nonparametric} data frame contains the bi-level selection
result.

\begin{CodeChunk}
\begin{CodeInput}
R> bamlasso_vs_part
\end{CodeInput}
\begin{CodeOutput}
$Parametric
character(0)

$`Non-parametric`
  Variable Linear Nonlinear
1       x1  FALSE      TRUE
2       x2  FALSE      TRUE
3       x3   TRUE     FALSE
4       x4  FALSE      TRUE
5       x5  FALSE      TRUE
6       x7  FALSE      TRUE
7       x9  FALSE      TRUE
8      x10  FALSE      TRUE
\end{CodeOutput}
\end{CodeChunk}

\subsubsection{Curve Plotting}

We also provide a utility function \texttt{plot\_smooth\_term} to plot
the estimated functions. The function takes in the fitted model, the
variable name, the previously constructed smooth objective to construct
the design matrix, minimum and maximum of the range of the predictors.
The function outputs a \texttt{ggplot} object to show the estimated
curve.

\begin{CodeChunk}
\begin{CodeInput}
R> plot_smooth_term(bham_final, "x1", train_smooth,
+                      min = min(dat[, "x1"]),
+                      max = max(dat[, "x1"]))
\end{CodeInput}
\begin{CodeOutput}
`geom_smooth()` using method = 'loess' and formula 'y ~ x'
\end{CodeOutput}


\begin{center}\includegraphics{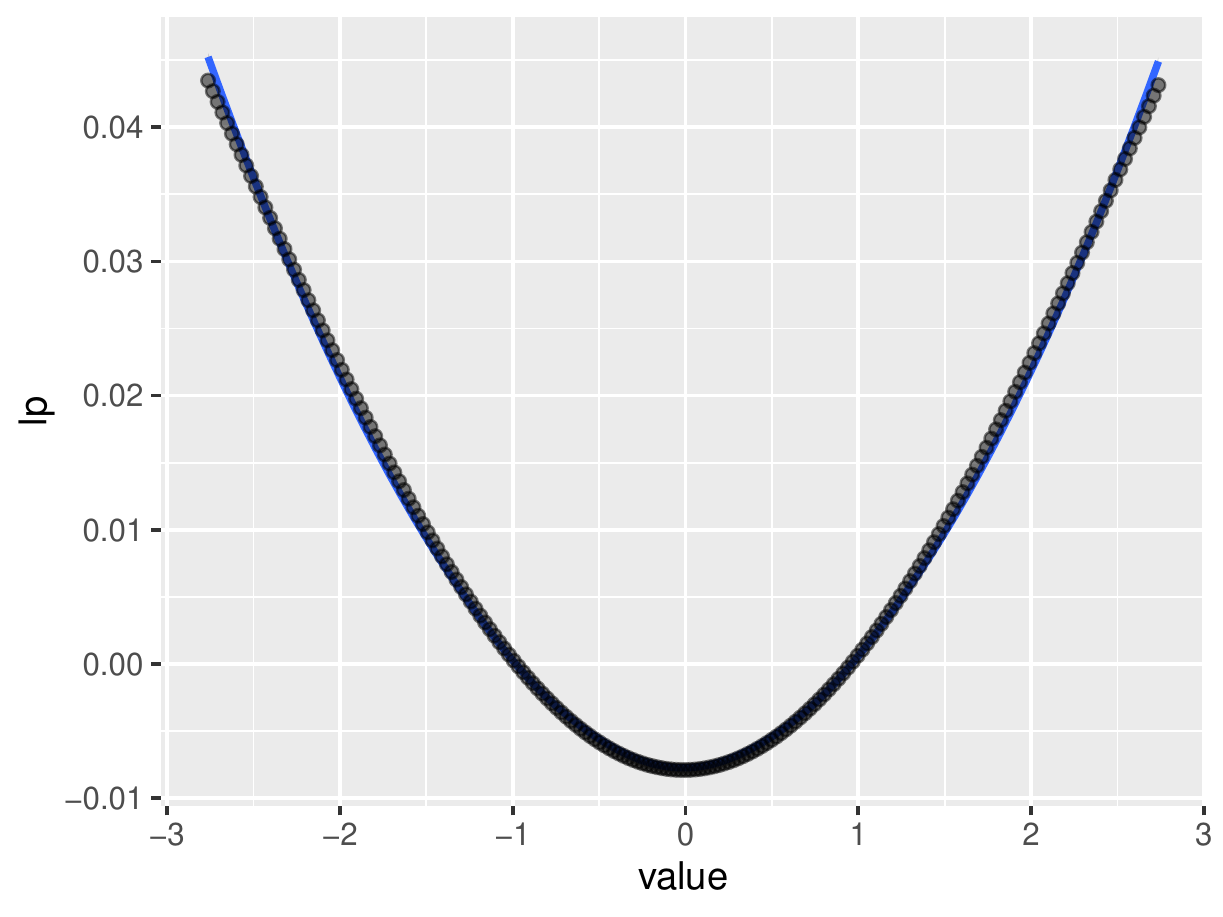} \end{center}

\end{CodeChunk}

\subsection{Prediction}

To predict new datasets, we need to go through the same two-step
procedure to produce the data matrix as previously when building the
model. First of all, we need to translate the new dataset to their
matrix form using the function \texttt{make\_predict\_dat}. This step is
necessary because of the reparameterization of the design matrix. The
function \texttt{make\_predict\_dat} is based on the function
\texttt{PredictMat} from \texttt{mgcv}. The function asks for an
additional input argument besides the new dataset, which is the Smooth
object when constructing the design matrix for the training data. The
output of the function is the new data matrix of the new dataset with
conformable dimension and variable name. We show the first six columns
of the first five observations in the following example.

\begin{CodeChunk}
\begin{CodeInput}
R> train_smooth <- train_sm_dat$Smooth
R> test_sm_dat <- make_predict_dat(train_sm_dat$Smooth, dat = test_dat)
\end{CodeInput}
\end{CodeChunk}

\begin{CodeChunk}
\begin{CodeOutput}
     x1.pen1     x1.pen2    x1.pen3     x1.pen4    x1.pen5   x1.null1
1  0.2105822 -0.51339049 -0.7087016  1.48599984 -1.6282845 -0.6187856
2  0.1401345 -0.04244866  0.4101826 -2.49600416 -0.5234851  1.2762868
3 -0.1157559 -0.28695472  0.3280699  0.08111065  9.0232660  3.6778234
4 -0.1551403 -0.57575563  1.1032603 -2.71644734  1.1193043  1.8444025
5  0.1841429 -0.50899800 -0.7593021  1.41065378 -1.6744694 -0.5829544
\end{CodeOutput}
\end{CodeChunk}

With the new dataset in the conformable design matrix format, we can
easily produce the prediction using the function \texttt{predict}. Under
the hood, we use \texttt{predict.glmnet} to produce the prediction, and
hence, it is robust. For the GLM, we can produce the linear predictors
using \texttt{type\ =\ "link"} and the fitted probability/mean using
\texttt{type\ =\ "response"}.

\begin{CodeChunk}
\begin{CodeInput}
R> bham_final$offset = 0
R> pred_res <- predict(bham_final, newx = as.matrix(test_sm_dat),
+                     newoffset = 0, type = "link")
\end{CodeInput}
\end{CodeChunk}

To note, we suggest to use \texttt{BhGLM::measure.bh} to provide a quick
prediction performance evaluation for the new dataset.

\begin{CodeChunk}
\begin{CodeInput}
R> if(!require("devtools")) install.packages("devtools")
R> if(!require("BhGLM")) devtools::install_github("nyiuab/BhGLM")
R>
R> BhGLM::measure.bh(bham_final, as.matrix(test_sm_dat), test_dat$y)
\end{CodeInput}
\end{CodeChunk}

\section{Discussion}

In this article, we introduce the R package \texttt{BHAM} to fit
Bayesian Hierarchical additive models with two-part spike-and-slab LASSO
prior. Specifically, \texttt{BHAM} provides a flexible and scalable
solution to fit high-dimensional generalized additive model and additive
Cox model for continuous, discrete and time-to-event outcomes. To help
users to familiarize \texttt{BHAM}, we demonstrate the analytic pipeline
in this manuscript. We illustrate additive model construction for
high-dimensional data, model fitting and tuning, signal selection and
visualization, and new data prediction. Our demonstrating analytic
pipeline for binary outcomes can be easily translate to other outcomes
by using different options in the functions. We recommend readers to
visit our interactive website\url{https://boyiguo1.github.io/BHAM/} for
more examples.

Our R package \texttt{BHAM} is versatile and can be widely applied in
large-scale molecular and clinical data analyses to model complex
signals and provide improved prediction accuracy. Compared to the
``black-box'' machine learning methods, the additive models provide more
interpretable inference of underlying signals, for example using simple
visual presentations. In addition, \texttt{BHAM} uniquely offers a
bi-level selection approach to detect if underlying signals are linear
or nonlinear. This signal detection procedure is a natural product of
the two-part spike-and-slab LASSO prior and requires no further
thresholding or hypothesis testing. This feature also grants flexibility
and automation to our models to relieve users from a priori assumptions
on underlying signals. In other words, users do not have to go through
any laborious tests for signals shape before fitting a model.
\texttt{BHAM} also innovates the computation aspect of high-dimensional
Bayesian additive model. It is widely known that fitting
high-dimensional Bayeisan models are computationally, particularly when
approximation algorithms are used. We provide an economic solution by
integrate coordinate descent algorithm with the EM procedure. Our
implementation leverages some commonly used modeling interface form the
standard R packages, for example \texttt{glmnet} for coordinate descent
algorithm, and hence guarantees robustness. Lastly, \texttt{BHAM} allows
full customization of additive functions and offers users more
flexibility to conduct analyses. We follow the popular additive function
syntax in \texttt{mgcv} and provide a parse function to easily produce
high-dimensional formula. This level of customization is rarely provided
in other high-dimensional additive model packages.

In conclusion, we offer an R package \texttt{BHAM} to provide a flexible
and scalable solution to model complex signals for high-dimensional
data. We aim to optimize the high-dimensional additive modelling
experience by providing friendly analytic pipeline, easy additive
function customization and fast algorithms. Our package provides
interpretable inference and improved prediction, and hence contribute to
prognostic research.

\clearpage

\bibliography{references.bib}

\end{document}